%% file: 0-IEEEtran_main.tex
\documentclass[lettersize,conference]{IEEEtran}
\usepackage{amsmath,amsfonts}
\usepackage{algorithmic}
\usepackage{array}
\usepackage[caption=false,font=normalsize,labelfont=sf,textfont=sf]{subfig}
\usepackage{textcomp}
\usepackage{stfloats}
\usepackage{url}
\usepackage{verbatim}
\usepackage{graphicx}
\usepackage{array, longtable, booktabs, makecell}
\usepackage[numbers]{natbib}
\def\BibTeX{{\rm B\kern-.05em{\sc i\kern-.025em b}\kern-.08em
    T\kern-.1667em\lower.7ex\hbox{E}\kern-.125emX}}
\usepackage{balance}

\usepackage{longtable}

\usepackage{array, longtable, booktabs, makecell}
\usepackage{placeins}

\newcounter{rownum}
\usepackage{url}

\usepackage{breakurl}

\usepackage{pdfpages}
\usepackage{todonotes}

\usepackage{amsfonts,bbm,bm,courier}
\usepackage{comment}

\usepackage{color,soul}

\newcommand{\x}{\bm{x}}

\begin{document}
\title{Explainable agency: human preferences for simple or complex explanations}

\author{
\IEEEauthorblockN{Michelle Blom\IEEEauthorrefmark{1}, Ronal Singh\IEEEauthorrefmark{3}, Tim Miller\IEEEauthorrefmark{4}, Liz Sonenberg\IEEEauthorrefmark{1}, \\ Kerry Trentelman\IEEEauthorrefmark{2}, Adam Saulwick\IEEEauthorrefmark{2}\\}
\IEEEauthorblockA{\IEEEauthorrefmark{1}School of Computing and Information Systems\\
       The University of Melbourne, Melbourne, Australia \\
       \{michelle.blom, l.sonenberg\}@unimelb.edu.au\\}
\IEEEauthorblockA{\IEEEauthorrefmark{2}Defence Science and Technology Group, Adelaide, Australia\\
       \{kerry.trentelman, adam.saulwick\}@defence.gov.au}\\
\IEEEauthorblockA{\IEEEauthorrefmark{3}CSIRO Data61,
       Melbourne, Australia \\
       ronal.singh@data61.csiro.au\\}
\IEEEauthorblockA{\IEEEauthorrefmark{4}School of Electrical Engineering and Computer Science,\\
    The University of Queensland, Brisbane, Australia \\
       timothy.miller@uq.edu.au}
}

\markboth{IEEE Transactions on Human Machine Systems,~Vol.~53, No.~4, August~2023}%
{}

\maketitle

\begin{abstract}
\input{01-abstract}
\end{abstract}


\input{1-introduction}
\input{2-lit}

\input{3-study-design}

\input{4-results}

\input{5-discussion}

\input{6-conclusion}

\bibliographystyle{IEEEtranN}
\bibliography{library}

\newpage

\input{7-appendix}

\end{document}

%% file: 01-abstract.tex
 Research in cognitive psychology has established that whether people prefer simpler explanations to complex ones is context dependent, but
the question of `simple vs.\ complex' becomes critical when an artificial agent seeks to explain its decisions or predictions to humans. 
We present a model for abstracting causal reasoning chains for the purpose of explanation. This model  uses a set of rules to progressively abstract different types of causal information in causal proof traces. We perform online studies using 123 Amazon MTurk participants and with five industry experts over two domains: maritime patrol and weather prediction. We found participants' satisfaction with generated explanations was based on the consistency of relationships among the causes (\emph{coherence}) that explain an event; and that the important question is not whether people prefer simple or complex explanations, but \emph{what} types of causal information are relevant to individuals in specific contexts.

%% file: 1-introduction.tex
\section{Introduction}
\label{sec:intro}

 Explanations for questions such as `Why is the fog so heavy this morning?' are not simple, despite the simple nature of the question. How much detail should be included in the explanation? How complex or how simple should we make an explanation, and which is more satisfactory? With the spike in algorithmic decision-making, explaining algorithmic decisions at the right level of complexity and abstraction has become increasingly important. 
 
In this paper, we have designed a system that explains proof traces generated by a first-order logic-based reasoning system -- Consensus \cite{lambert_consensus_2015} -- which is designed to automatically detect events, for example, whether a marine vessel is engaged in illegal fishing~\cite{trentelman_information_2019}. Consensus takes the information of varying modalities from different sources, such as Automated Information Systems (AIS) vessel movement data and textual data written in controlled natural language to reason over and infer behaviours and relationships~\cite{hutchison_lexpresso_2014}. While the proof traces generated may be suitable for a knowledge engineer to debug system behaviour, the complexity of the proof traces is such that other potential users, for example, intelligence system operators, cannot easily interpret them as input to informed decision-making. Therefore, explaining algorithmic decisions at the right level of complexity and abstraction is important.  

Based on existing methods of summarising and simplifying proof traces~\cite{carbonell_reconstructing_1994,horacek1999presenting,horacek_how_2007,furtado_abstracting_2007}, we have designed three rules to generate explanations at three levels of complexity. Applying fewer (or more) rules results in a complex (or simple) explanation, where each rule removes a particular type of causal information. Our explanation system takes proof traces generated by Consensus and progressively applies these rules in a specific order to increasingly generate simpler explanations. Note that all information in an explanation is causally relevant to the Consensus reasoning system. 

There are many ways to characterise the complexity or simplicity of an explanation, and there is no consensus on any one definition~\cite{lombrozo2017causal,zemla2017evaluating}. We adopt \textit{node simplicity} of \citet{pacer2017ockham} to measure explanation complexity by counting the number of causes in the explanation. A simpler (or more abstract) explanation is one that has fewer causes. 

Based on existing methods of summarising and simplifying proof traces~\cite{carbonell_reconstructing_1994,horacek1999presenting,horacek_how_2007,furtado_abstracting_2007}, we have designed three rules to generate explanations at three levels of complexity.  
 Applying fewer (or more) rules results in a complex (or simple) explanation, where each rule removes a particular type of causal information. 
 Our reasoning system takes proof traces generated by Consensus and progressively applies these rules in a specific order to increasingly generate simpler explanations. Note that all information present in an explanation is causally relevant to the reasoning system.

 We conducted a study with 123 participants from the general public and five industry experts with a background in the application domain. Since the three rules remove causal information to generate a more abstract explanation, we first aimed to identify when to remove or retain a given type of causal information. We wanted to understand what causal information types are important for the simplest explanation of an event. Our second aim was to verify whether people prefer simple or complex explanations generated by our system and how well this aligns with existing literature.

 We observed a preference for complex explanations, which is consistent with existing findings~\cite{johnson2019simplicity,zemla2017evaluating}; however, we also found some simplification was necessary, meaning that we needed to apply at least one rule to remove low-level logical reasoning. This suggests that while all information was causally relevant to the reasoning system, people did not need or want it. Participants rated explanations based on the consistency of relationships among the causes; that is, explanation coherency~\cite{thagard1978best,bovens2000coherentism,bovens2003solving,koslowski2008information,zemla2017evaluating}. In our case, this translates to not applying rules that remove information promoting explanation coherency. Furthermore, the role of causal information impacts the preference for complexity. Participants rated the more complex explanation lower if it had causes that were assumed common sense knowledge or unnecessary background knowledge. It was unclear when to apply or not apply rules that remove such information ---  these were scenarios and individual dependent (experts did not need the background knowledge). It suggests that we need to investigate this finding further. \\

The remainder of the paper is organised as follows. In Section~\ref{sec:litreview}, we discuss relevant literature.  Section~\ref{sec:SystemDescription} discusses the design of our explanation system. In Section~\ref{sec:StudyDesign}, we discuss our study design and in Section~\ref{sec:results}, our results. We discuss key findings, limitations and future work in Section~\ref{sec:discussion}.  

%% file: 2-lit.tex
\section{Background}
\label{sec:litreview}

Consensus supports where, what, when, and who based queries \cite{lambert_consensus_2015,lambert_mephisto_2008}. A person may ask ``Who was illegally fishing?''. Such queries are translated into logical formulae that Consensus will try and satisfy \cite{trentelman_information_2019}. Queries are answered using a Boolean satisfiability (SAT) solver in conjunction with all available knowledge across its memory. For the conclusion ``The Sea Witch was illegally fishing'', Consensus can generate a proof trace, describing the inference steps necessary to infer the associated logical formulae. We present, in this paper, a system to transform these proofs in order to provide explanations to humans at varying levels of abstraction.

\subsection{Explaining Proofs and Proof-based Reasoning}\label{sec:LogicBased}

Machine generated proofs can be challenging to interpret by a human user, even if that user is a mathematician or an expert in automated theorem proving. There has been much work, dating back to the 1970s, on the simplification of formal proofs, and the translation of proofs into text. The degree of simplification achieved varies, with early work focused more on direct translation. Later work considers proof abstraction. The written translations of these simplified proofs are still targeted at logicians or mathematicians. While we do not exhaustively examine the body of work on proof reformulation and translation, the systems described below are reflective of the range of approaches developed and their evolution.

\citet{horacek1999presenting,horacek_how_2007} presents a five step process for selecting what to explain when forming an explanation of a proof. Given a proof trace in some calculus, the proof is translated into a more `human-oriented' calculus. Proof structures, expressed as a tree, are then shrunk to their essential arguments, taking into account user knowledge. Parts of the proof that are deemed to represent cognitively challenging patterns of reasoning are expanded to finer grained inferences. The result is a proof at a partial assertion level with mixed granularities. The final steps remove selected elements and reorganise inferences into a coherent structure. Trivial premises and inference steps, and references to axioms that can be inferred from a statement of the premises and resulting conclusion, are omitted.

\subsubsection{Explanation and the Semantic Web}

Semantic Web applications often use logic-based methods to represent and reason with information. Typical Semantic Web applications use heterogeneous sources of information, of varying provenance, to make decisions or form conclusions. \citet{mcguinness_explaining_2004} describe the Inference Web (IW), a prominent infrastructure for reasoning in the Semantic Web. The IW makes use of proof abstraction methods to reformulate proofs into a form `more amenable for human consumption'. This is achieved through custom proof rewriting rules. \citet{mcguinness_explaining_2004}, and other IW publications, do not elaborate on the  rewriting rules used. Generally speaking, these rules are likely based on omitting intermediate inferences, hiding core reasoner rules,  abstracting proofs to the assertion level as described by \citet{carbonell_reconstructing_1994}, and using custom abstraction patterns \cite{furtado_abstracting_2007}. An explanation can be viewed as a layered `composite object' that can be interrogated for the appropriate level of detail in any context \cite{preece_stakeholders_2018}.

More recent work on inference in a Semantic Web context has focused on the construction, and querying, of knowledge graphs.    
Graph abstraction methods are typically based on finding clusters of nodes that can be merged, or patterns of inference that can be replaced.  \citet{missier2013provenance}  define a mapping between specific patterns that may appear in a provenance graph and simpler patterns to replace them. \citet{missier2014provabs} and  \citet{moreau2015aggregation} focus on grouping nodes according to their `provenance type'. \citet{chen2012visualization} combine user directed aggregation of nodes with rules for replacing specific relationship patterns.  
Most work on graph abstraction is designed to preserve the \textit{meaning} or validity of a graph -- ensuring that paths that were present in the original graph are still present, and no new paths are created.

 \subsection{User perception of complexity in explanations}
 \label{sec:lit-userstudies}
 
Several theories exist on whether and when people prefer complex or simple explanations. \citet{lombrozo2007simplicity} performed studies using scenarios designed on alien diseases and their potential causes and found that participants generally favoured simpler explanations, but they chose more probable ones when probability information was available, even when the explanation was more complex. When people initially favour simpler explanations, presenting strong, specific evidence that aligns with a more complex explanation can change their preference.

\citet{zemla2017evaluating} assessed the role of several explanatory criteria in predicting judgements of explanation quality. They conducted two studies using explanations from Reddit's Explain Like I'm Five (ELI5\footnote{www.reddit.com/r/explainlikeimfive}), Wikipedia, and HowThingsWork.com. To assess the simplicity of an explanation, they counted the number of causal mechanisms in the causal models that the experimenters created from the explanations, the explanation length and the explanation language. They found that more complex explanations were more highly rated. The number of causal mechanisms and explanation length both predicted the explanation quality. They also indicated that the perceived expertise of the explainer impacted the quality positively. People prefer complex explanations that involve more causal mechanisms to explain an effect. This preference for complexity is attributed to a desire to identify a sufficient number of causes to make the effect seem inevitable.
 
The opponent heuristic theory of \citet{johnson2019simplicity} states that usually, simplicity is a good cue for an explanation's prior probability, $P(H)$, which is consistent with Lombrozo's findings~\cite{lombrozo2007simplicity}; while complexity is a good cue for an explanation's likelihood or fit to the evidence, $P(E|H)$, since complex causes have more opportunities to cause each aspect of the evidence. In their studies, the probability of causes was included in the explanations, and how this affected participants' preferences was tested. Participants showed a preference for complex explanations in stochastic contexts. Participants preferred simple explanations to differing degrees across domains; participants favoured simple explanations to a greater degree for physical over social systems. This nuanced approach allows for a dynamic adjustment of explanatory preferences, demonstrating that neither complexity nor simplicity is universally preferred. Rather, preference depends on the specific context and the nature of the evaluated explanation.

\citet{Lim2020} propose the complexity matching hypothesis---that people believe a satisfying explanation should be as complex as the event being explained. They designed scenarios with a varying number of relevant details. Complex versions had three details describing the event, while simple versions had only one detail. They found a general preference for more complexity (as explanatory complexity increases, satisfaction increases for simple and complex scenarios), and a preference for complexity matching (satisfaction increases at a higher rate for complex scenarios).  
 
 Additionally, there has been a discussion on whether causally relevant information makes a difference in the preference for complexity. Our studies do not intentionally add causally irrelevant information to explanations. All information presented in our explanations is causal to the reasoning system. Our user evaluation of these explanations aims to understand which types of causal information are relevant for people to understand the prediction being explained and, therefore, what to include to make explanations simple or complex. We also note that our investigation is about an artificial agent explaining its decisions to a human.

 \citet{bechlivanidis2017concreteness} conducted experiments to investigate whether people preferred concrete or abstract explanations. They investigated three types of explanations that differed in their degree of concreteness: concrete, abstract, and irrelevant. The first two explanations (concrete and abstract) had three key facts, and the facts were abstracted in the abstract explanation. Three other facts were included only in the irrelevant explanation and omitted from the other two. 
 A preference for detail was discovered, even when the details were considered causally irrelevant to the explanandum. Moreover, the participants did not penalise irrelevant factors or details with minimal causal relation to the explanandum. However, they did penalise concrete explanations with concrete terms that failed to communicate an event's causal properties.

\citet{weatherson2012explanation} introduces the idea that good explanations strike a balance between too much and too little detail, which has subsequently been echoed by others~\cite{burgoon2013there,Boone2016-ff,craver2020more}. Unlike overly specific explanations, which may fail to adequately capture the broader context or miss critical interventions, moderately abstract explanations provide a balanced approach that is precise in identifying relevant interventions and avoids unnecessary detail that could obscure the explanation's utility.

\citet{aronowitz2020experiential} note that most people answer why questions with experiential explanations -- `narratives or stories with a temporal structure and concrete details'. Abstracted explanations, in contrast, incorporate general facts and fewer details. \citet{aronowitz2020experiential} argue that while abstract explanations have value, particularly in their ability to support generalisation, they are not always preferred to those with greater detail, and experiential explanations have several advantages. Experiential explanations provide sufficient detail and context to be used as inputs for `mental simulation', aiding the recipient in understanding a scenario, and can be `repurposed' in new contexts.  
 
\citet{blanchard2020explanatory} argue that people prefer explanations that strike the right balance between `specificity and generality'. Explanations that are too abstract do not allow recipients to identify specific `interventions that would have changed the explanandum', while details that are not useful for identifying or distinguishing between different interventions should be omitted. They regard moderately abstract explanations as optimal, because they are specific, and at the same time abstract enough, to identify explanandum-changing interventions.

More generally, we have to deal with `families of abstractions'~\cite{craver2020more}, each with different degrees of abstraction and simplifications or idealisation~\cite{Boone2016-ff}. These abstractions work collaboratively to help us better understand a phenomenon, suggesting that the desire for more detailed and abstract explanations is driven by the \textit{level} at which we want to understand a phenomenon.

\noindent\textbf{Summary}
This discussion indicates a nuanced preference for simple versus complex explanations. Preferences oscillate between simplicity and complexity based on context, evidence, and the explanations' ability to convey causally relevant information effectively. The presence of causally relevant information significantly influences preference towards complexity.

%% file: 3-study-design.tex
\section{Problem Definition}\label{sec:ProblemDefinition}

Given a \textit{proof trace}, $T$, to establish the truth of a conclusion $c$ using a rule-based system of logical reasoning, our goal is to form a layered explanation to answer the question `\textit{Why} $c$?'. Each layer contains a natural language description of \textit{why} $c$ was inferred by the reasoning system, with consecutive layers describing progressively fewer inferences, premises, or intermediate conclusions of the reasoning process. 

The design of this system is described in Section \ref{sec:SystemDescription}. At a high level, the system extracts specific types of knowledge from a trace (Section \ref{sec:KnowledgeTypes}) and uses this knowledge to form a graphical representation -- an explanation graph ($\mathcal{G}$) -- of the reasoning documented in the trace such that $\mathcal{G} \models c$  (Section \ref{sec:FormProofGraph}). A set of three simplification rules are progressively applied to this graph to form a sequence of increasingly more abstract explanation graphs ($\mathcal{G'}$) such that $\mathcal{G'} \models c$, that is, the user can reach the same conclusion ($c$) using either $\mathcal{G}$ or $\mathcal{G'}$  (Section \ref{sec:SimplifyingRules}). Therefore, this aims to abstract the graph such that the abstraction preserves the ability to infer why $c$; e.g. removing `common' knowledge or redundant knowledge. Each of these graphs is translated into  natural language to form a layer of the produced explanation\footnote{For brevity, we omit discussion of the process used to translate an explanation graph into natural language in this paper.}. 

\section{Explanation System}\label{sec:SystemDescription}
\begin{figure*}[t]
\centering
\includegraphics[width=0.55\textwidth]{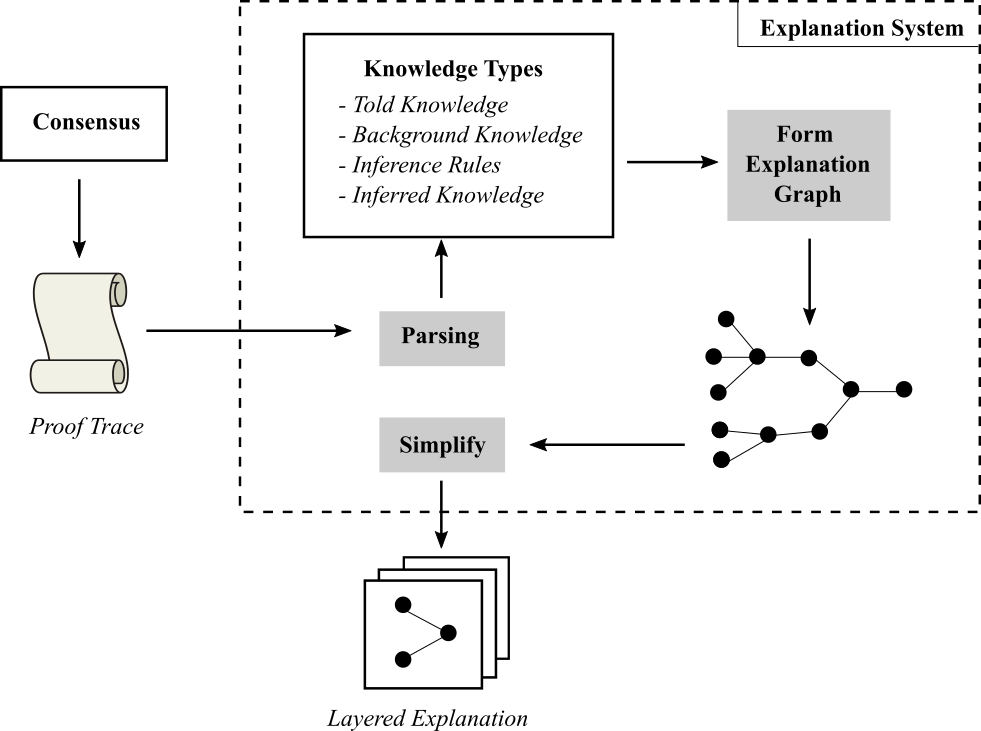}    
\caption{Process employed by our explanation system to produce a layered explanation from a proof trace. The proof trace is parsed to extract different kinds of knowledge (Section \ref{sec:KnowledgeTypes}), which is then used to form an explanation graph (Section \ref{sec:FormProofGraph}). The graph is simplified via the application of a number of rules (Section \ref{sec:SimplifyingRules}) to form a layered explanation.}
\label{fig:Process}
\end{figure*}

Figure \ref{fig:Process} depicts the design of our explanation system. The system takes a \textit{proof trace}, produced by Consensus in notation modelled on PROV \cite{groth2013prov,missier2018provenance}, parses the trace to extract key types of knowledge, and uses that knowledge to form a graphical representation of the reasoning used to arrive at a conclusion. The language used to encode premises and conclusions in these traces can be varied, with the explanation system itself not dependent on any one language in particular. 

\subsection{Knowledge Types}
\label{sec:KnowledgeTypes}
In parsing a proof trace, we collect the following types of knowledge: 
\begin{enumerate}
    \item 
        \textbf{Told Knowledge:} knowledge that was told to Consensus by a human user. For example, a user might tell the system--in controlled natural language--that they are at a given location at a certain time.

    \item 
        \textbf{Background Knowledge:} prior knowledge that the system has about a domain, for example: the location of air and sea ports; air and sea lane waypoints; and the boundaries of Exclusive Economic Zones.

    \item 
        \textbf{Inference Rules:} logical rules, such as \emph{modus ponens} and conjunction introduction/elimination, and semantic rules  from axioms and definitions. Each rule connects one or more premises to a conclusion.  For example, if a physical thing $A$ is connected (as per the Region Connection Calculus \cite{cohn1997rcc}) to a physical thing $B$, and $B$ is at a geographic coordinate $C$, then $A$ is also at $C$.

    \item 
       \textbf{Inferred Knowledge:} knowledge inferred by the reasoning system after applying \emph{Inference Rules}. 
\end{enumerate}
We form a graphical representation--an \textit{explanation graph}--of the reasoning undertaken by Consensus in the trace. This graph is \textit{simplified} via an ordered application of a number of \textit{rules}, to generate explanations at varying levels of simplicity (or complexity). Each level or \textit{layer} of the explanation corresponds to the application of differing numbers of simplifying rules.

\subsection{Explanation Graphs}
\label{sec:FormProofGraph}
An explanation graph is constructed from the four types of  knowledge we extract from a proof trace. The graph is designed to be a faithful representation of the reasoning that has occurred in the trace. We define an explanation graph $\mathcal{G}$ in terms of a collection of nodes $\mathcal{N}$ and edges $\mathcal{E}$, $G = \langle \mathcal{N}, \mathcal{E}\rangle$. Each node $n \in \mathcal{N}$ is  
a piece of content of one of the above four knowledge types.
Using the  knowledge extracted from a proof trace, an explanation graph is constructed as follows: 
\begin{enumerate}
    \item A node is created for each piece of  Told and Background Knowledge, and connected to its associated piece of Inferred Knowledge (eg. $T_i \rightarrow I_j$). For example, we may be told that Paula Lands is on a yacht in the Pacific Ocean ($T_i$). Here, $I_j$ is the location of Paula Lands. 
    
    \item A node is created for each additional piece of Inferred Knowledge.
    
    \item For each Inference Rule application, we create a node representing that rule application in our graph. The knowledge used as premises are connected to our rule node (i.e., $I_i \rightarrow R_j$), with the rule node then connected to the conclusion being inferred (eg. $I_i \rightarrow R_j \rightarrow I_k$).  
\end{enumerate}

Nodes without parents represent root causes. Edges represent causal relationships between these information types.  Told and Background Knowledge represent the root causes in the explanation graphs generated by our explanation system. 

\subsection{Simplifying Rules}\label{sec:SimplifyingRules}

Recall that $\mathcal{G} \models c$, where $\mathcal{G}$ is the explanation graph and $c$ is the conclusion. Our explanation system simplifies an explanation graph, $\mathcal{G}$, with the application of a number of rules to generate explanations at varying levels of simplicity (or complexity) such that $\mathcal{G'} \models c$, where $\mathcal{G'}$ is the abstract explanation graph after applying one or more rules. Our goal is for the user to still reach the conclusion $c$ with the information included in the graph $\mathcal{G'}$. 
The explanation system generates explanations at varying levels of simplicity (or complexity), each a \emph{layer} of a \emph{layered explanation}.  
Each new layer is formed by applying a rule to the layer below. 

\begin{enumerate}
    \item 
        \textbf{Flatten Logic (FL)} Existing proof abstraction techniques remove reference to axioms or steps that can be inferred from simply stating the premises and resulting conclusion \cite{carbonell_reconstructing_1994,horacek1999presenting,horacek_how_2007,furtado_abstracting_2007}. This is applied in our context by flattening the application of the conjunction introduction logic rule, and removing the `restatement' of told knowledge (i.e., we are told $X$ and therefore $X$ holds).  
        For all [$P_1$,$...$,$P_n$] $\xrightarrow{}$ Rule[Conjunction] $\xrightarrow{}$ $I$, where $P_i$ is any type of knowledge, and $I$ is a conclusion are flattened to [$P_1$,$...$, $P_n$] $\xrightarrow{}$ $I.$  We substitute $T$ for all   $T$ $\xrightarrow{}$ $I$, where $T$ is a piece of Told or Background Knowledge, and $I$ is a restatement of that knowledge. 
  
    \item 
        \textbf{Flatten Rules (FR)} Omit rule nodes that connect one or more inferred premises to a conclusion or intermediate conclusion. This rule assumes, as described above, that the inferential step connecting this collection of nodes can be inferred from a statement of the premises and conclusion. 
        All [$P_1$,$...$, $P_n$] $\xrightarrow{}$ $R$ $\xrightarrow{}$ $I$, where $R$ is the application of a rule, are flattened to [$P_1$,$...$, $P_n$] $\xrightarrow{}$ $I$. 
        
    \item 
        \textbf{Filter Knowledge (FK)} Remove nodes from the graph that are considered to contain `less relevant' or background knowledge. This step reflects the selection phase of proof simplification expressed by \cite{horacek1999presenting,horacek_how_2007}. Background knowledge is usually not representative of a behaviour or action undertaken by an entity in the domain, but of contextual or general knowledge assumed by the Consensus user. We connect the parents of a removed node $n$ to the children of $n$ by a directed edge.
\end{enumerate}

These three rules are based on typical translations present in existing work on proof reformulation, as discussed in Section \ref{sec:LogicBased}, and are likely to generalise to other kinds of logical proof traces beyond those produced by Consensus. The most complex or detailed explanation is one that has had no rules applied to it. The first level of abstraction applies only the \textit{Flatten Logic} rule, while the simplest explanation results after applying all three rules, \textit{Flatten Logic} $\rightarrow$ \textit{Flatten Rules} $\rightarrow$ \textit{Filter Knowledge}.
 It is most natural to apply Flatten Logic first, as this removes information that is least likely to have explanatory value.  We consider the relative value of the final two rules, in terms of whether they result in more preferable explanations, as part of our study (Section \ref{sec:StudyDesign}) where  we compare explanations formed by different rule combinations.

\section{Study Design}\label{sec:StudyDesign}
Our study uses two domains --  weather events and illegal activity at sea (maritime). In the weather domain, explanations are provided for various weather or natural events such as cyclones, excessive fog, and volcanic eruption. In the maritime domain, scenarios involve various kinds of illegal activity at sea or sea ports, such as fishing in protected areas or regions without an appropriate licence.  We designed 18 scenarios; 11 from the maritime domain and 7 from the weather domain. 

 The MTurk study was formed by presenting a participant with series of pages, where on each page a \emph{conclusion}\footnote{We informed the participants that this was the system's \emph{prediction}.} was identified followed by two explanations side-by-side. The first explanation, denoted Explanation 1, was more abstract than the second, denoted Explanation 2\footnote{We ran a shorter study with 35 MTurk participants and confirmed that the order in which the explanations were presented (whether Explanation 1 was simple or complex) did not influence the preference for complexity.}. This pair of explanations compares two rule combinations, such as \emph{FL-FR} (simpler explanation) and \emph{FL} (more complex). Explanations were presented in natural language, as done in comparative works~\cite{lombrozo2007simplicity,zemla2017evaluating,johnson2019simplicity,Lim2020}. We refer the reader to  Appendix B  for a list of all pages considered in the study. With each page, we presented five questions on a seven-point Likert scale adapted from the explanation satisfaction scale by~\citet{hoffman_metrics_2018}. The ratings were compulsory. 
We provided an open-ended text response area for participants to allow additional optional feedback. We asked participants a \emph{Yes/No} question on whether they found Explanation 2 to contain more information than Explanation 1, such that the additional information aided in understanding the conclusion, and to justify their answer. The final question was compulsory. 

To keep the study manageable, we tested every layer with only the two layers below. For example, the \emph{no abstraction} layer (no rules applied) was compared with explanations generated by applying \emph{FL} and \emph{FL-FR}. Similarly, the explanation generated using the \emph{FL} rules was compared with explanations generated by applying \emph{FL-FR} and \emph{FL-FR-FK}.

\begin{table}[t]
\centering
\begin{tabular}{lc|lc}
\toprule
\textsc{Comparison} & \textsc{Set} & \textsc{Comparison} & \textsc{Set}\\
\midrule
FL vs none    & 1 &   FL-FK vs none & 2 \\[5pt]
FL-FR vs none & 1 & FL-FK vs none & 2 \\[5pt]
FL-FR vs FL             & 1 & FL-FK vs FL  & 2  \\[5pt]
\cline{3-4}
FL-FR-FK vs FL          & 1 & FL-FR-FK vs FL-FK       & 3  \\[5pt]
FL-FR-FK vs FL-FR       & 1 & FL-FR vs FL-FK          & 3 \\
\bottomrule
\end{tabular}
\caption{Rule combination comparisons evaluated through our study, involving rules  FL (Flatten Logic), FR (Flatten Rules), and FK (Filter Knowledge). We include the set to which they belong (Set 1--FR, Set 2--no-FR, and Set 3--FR vs no-FR).}
\label{tab:AllCombos}
\end{table}

The set of pair types (rule combination comparisons) were divided into three labelled sets.  The first set contains  explanation layer comparisons where the \emph{FR} rule has been applied. We call this category \emph{Set 1 (FR)}. The second set compares explanation layers in which the \emph{FR} rule has not been applied. This set is denoted \emph{Set 2 (No FR)}. Where \emph{FR} was not applied, the inference rules that we left in an explanation were included as footnotes in the explanation (this was a design decision). We also compared layers generated by the rule combination \emph{FL-FK} with layers generated by the rule combinations \emph{FL-FR} and \emph{FL-FR-FK}. This set of comparisons is denoted \emph{Set 3 (FR vs No FR)}. The objective of the third set was to learn the effect of keeping inference rule applications within an explanation. 
Table \ref{tab:AllCombos} lists all layer comparisons we consider in our study.

To remove scenario effects, we designed two pages for each of the pair types of interest, extracted from two different scenarios. For example, for the pair type \emph{FL vs no abstraction}, one instance of this pair was formed from an explanation predicting a blizzard and the second from an explanation on the likelihood of illegal fishing activities. Each participant was exposed to 6 randomly assigned pages was not presented with more than one page drawn from each scenario, or more than one page of each pair type. 

\subsection{Procedure}
\label{sec:study-procedure}
We conducted two studies, one through Amazon MTurk (Study 1), and a second study (Study 2) with five industry experts with a defence background.

Our first study was designed and administered as a Qualtrics\footnote{\url{https://www.qualtrics.com/au/?rid=ip\&prevsite=en\&newsite=au\&geo=AU\&geomatch=au}} survey. Before the experiment, we received ethics approval from our institution. Participants were paid USD \$14 per hour and a total of 123 people participated in the study. The participant played the role of an analyst using the reasoning system to make a critical decision. To help the analyst, the system presented two explanations. The analyst's role was to inform their team when they believed the system was making an important conclusion.  Participants received a plain language statement, a consent form, participated in a bot-check, and filled out a demographics questionnaire. Participants were  given a tutorial explaining the components of each page and what they  were expected to do when presented with a page.

\begin{table}
\scalebox{0.9}{
\begin{tabular}{ll}
\toprule
1 & From the explanation, I understand why the prediction has been made.\\
2 & The explanation of why the prediction was made provides sufficient detail.\\
3 & The explanation of why the prediction was made is satisfying.\\
4 & The explanation of why the prediction was made is complete.\\
5 & The explanation of why the prediction was made is trustworthy.  \\
\bottomrule
\end{tabular}}
\caption{Five questions, taken from taken from the satisfaction scale of \cite{hoffman_metrics_2018}, asked of participants to rate explanations.}
\label{tab:5Qs}
\end{table}

Each participant had 30 minutes to rate 6 side-by-side explanation pairs. Pages were presented sequentially without the option of going back and changing previous answers. 
Participants rated presented explanations using a seven-point Likert scale. The five questions forming the rating scale were taken from the satisfaction scale of \citet{hoffman_metrics_2018}, and are presented in Table \ref{tab:5Qs}. After rating explanations, participants were asked three further questions:
\begin{enumerate}
    \item \textit{Do you have any additional feedback regarding the ratings?}
    \item \textit{Explanation 2 contains information, not present in Explanation 1, that is helpful for understanding why the prediction was made}, with options  \textit{Yes}, \textit{No}, \textit{I don't know}.
    \item \textit{Please justify your answer to the previous question}. Participants were encouraged to provide a detailed answer. We asked this question to understand what participants were thinking and to judge whether the participants were appropriately engaged in the task. 
\end{enumerate}

For our second study, we recruited 5 participants who had a defence background, but were not necessarily working in the maritime domain. We ran the same study outlined above using only the 11 maritime scenarios. We presented 6 pages based on these scenarios at random to these defence experts.

%% file: 4-results.tex
\section{Results}
\label{sec:results}

Before performing any analysis, we manually reviewed the 123 answers of the Amazon MTurk participants to the last open-ended question. We removed those participants who provided nonsensical answers, or completed the overall task too quickly (e.g., 6 mins) or slowly (e.g., more than 30 mins). After this process, we were left with 111 participants, and all results in the subsequent discussions are based on these remaining participants. The mean task completion time was 20 minutes $\small(SD=5.3, min=10.2, max=29.4\small)$. 

\subsection{Data Analysis}
\label{sec:results-dataanalysis}
For each pair (rule combination comparison) of interest, we combined the ratings of the two groups and then performed tests by question. Each pair was rated by between 33 and 39 participants (note the different counts are due to removal of a few participants). The ratings provided were on a scale of \textit{Strongly Disagree} (1) to \textit{Strongly Agree} (7) with \textit{Neither Agree or Disagree} being option (4). The Friedman test was used to test whether the rating of Explanation 1 was significantly different from the rating of Explanation 2. All statistical testing was done using R, with a significance level of $0.1$. 

\subsection{Study 1: MTurk Participants}
\label{sec:results-study1}

We first analysed the participants' ratings for each question separately, and the \textit{difference} between the numerical ratings provided for Explanations 1 (more abstract) and 2 (more complex), denoted Ex1 - Ex2. 

\begin{figure*}[t]
    \centering
    \includegraphics[width=0.85\textwidth]{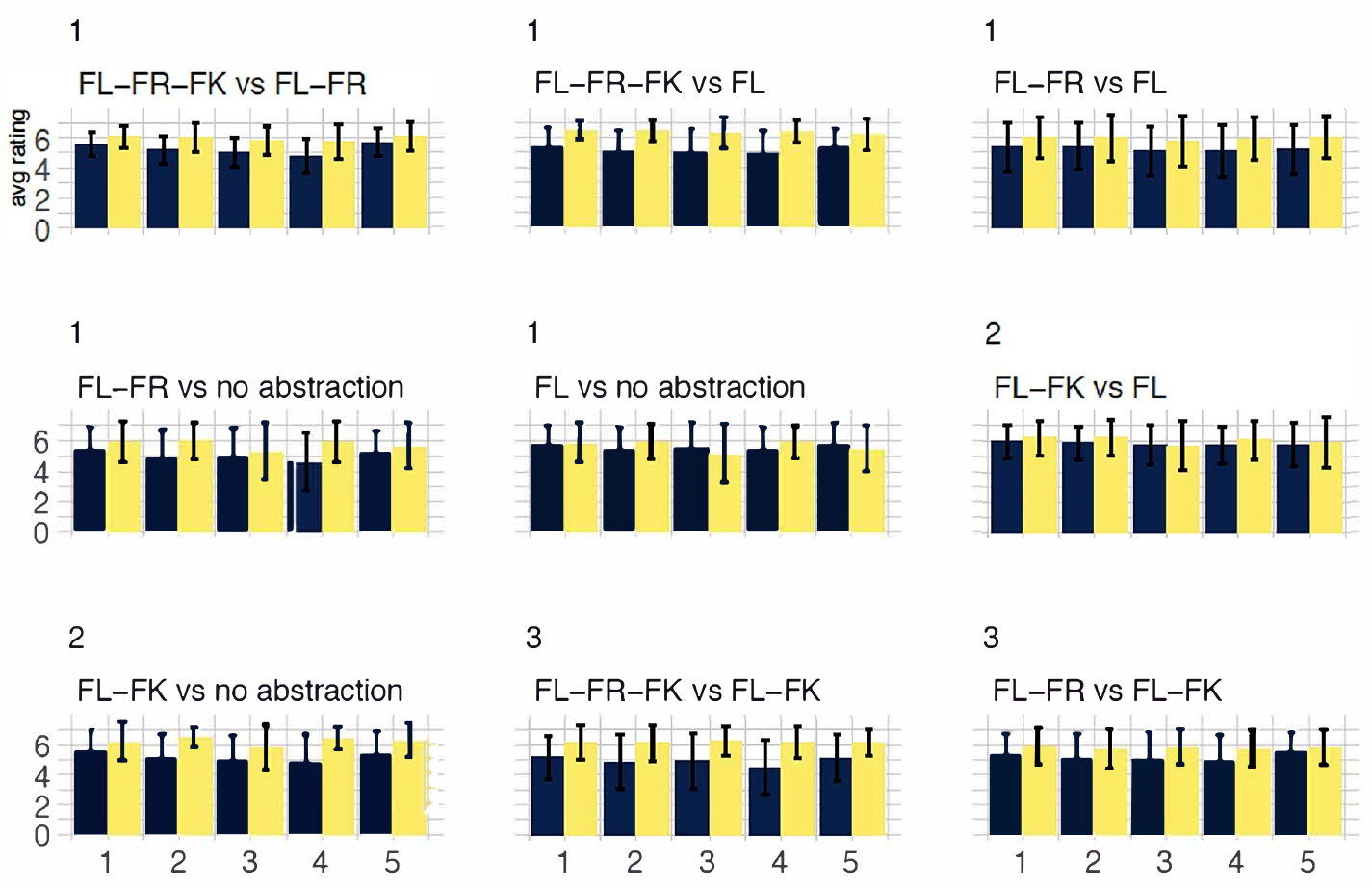}
    \caption{Average rating by question for rule combination comparison pairs. The bars on the left (dark blue) are  ratings for Explanation 1. The bars on the right (yellow) are ratings for Explanation 2. The title shows  Set number and pair type: Set 1 (FR), Set 2 (non-FR), and Set 3 (FR vs non-FR). The y-axis shows the mean rating for each question, with question numbers on the x-axis. Error bars denote one standard deviation. The five questions are outlined in Table \ref{tab:5Qs}.} 
    \label{fig:avg_by_pairs_by_question_2022}
\end{figure*}

Figure~\ref{fig:avg_by_pairs_by_question_2022} shows the average rating given to the two explanations by question. We observe that participants generally rated Explanation 2 (more complex) higher than Explanation 1 (more abstract) for each of the five questions. Moreover, participants generally found Explanation 2 to contain more information. 
While this general trend aligns with existing findings~\cite{zemla2017evaluating,johnson2019simplicity,Lim2020}, we highlight below some instances where this was not the case.

We performed the Friedman test, question-wise, to check whether the ratings for each pair were statistically different. See Appendix A for results. We found a statistically significant difference between the ratings across all five questions for 10 of the 18 pairs in which the participants rated the more complex explanations  higher. For \emph{FL-FR vs no abstraction} we did not find any significant difference between ratings for one of the five questions, which is: \textit{The explanation of why the prediction was made is satisfying.}  ($\chi^2\left(1\right) = 3.0, p = 0.08, Kendall's~W=0.44$). 

For three of the pairs, namely \emph{FL-FK vs FL}, \emph{FL-FR vs FL-FK}, and \emph{FL vs no abstraction}, there was no significant difference between the ratings across all five questions. We will discuss each of these pairs next.

\begin{figure*}[t]
    \centering
    \includegraphics[width=\textwidth]{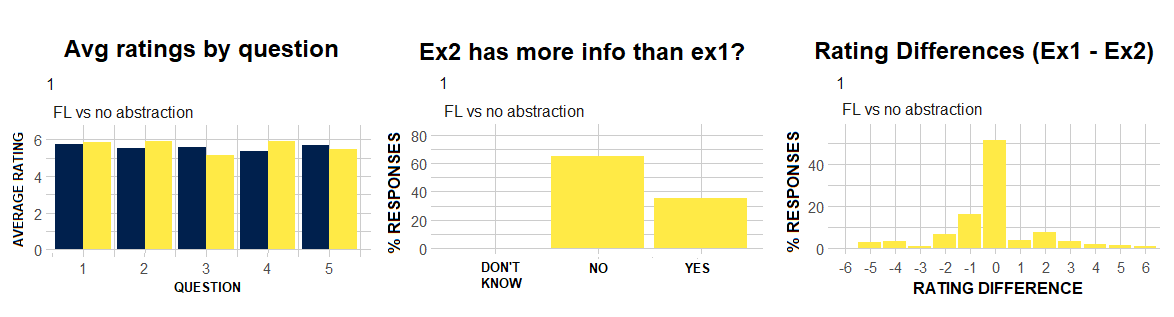}
    \caption{\emph{FL} vs \emph{no abstraction}. The first graph shows the rating for each of the five questions outlined in Table \ref{tab:5Qs}.  The second graph shows whether participants found Explanation 2 to contain more information than Explanation 1. The last graph shows the difference between the ratings given to the two explanations.}
    \label{fig:cf_noabs}
\end{figure*}

\textbf{FL vs no abstraction} \emph{FL} flattens logical conjunction and removes the `restatement' of told knowledge, while \emph{no abstraction} means that no rules were applied, generating the most complex explanation possible by our model. 
We observed no significant difference between the ratings of the two explanations across all five questions for this pair. We also see from Figure~\ref{fig:cf_noabs} (centre) that for this pair, 65\% of the participants stated that Explanation 2 did not contain any additional information (note that 65\% of the participants means that if 40 participants had rated this explanation, 26 supported the claim above). A qualitative analysis of the feedback shows that many participants found Explanation 2 to be repetitive, contain redundant information, or the same information as Explanation 1.   
Through the use of \emph{FL}, the model generated explanations that removed some of the reasoning information relevant to the system, but perhaps was not that useful to the participants.

\begin{figure*}[t]
    \centering
    \includegraphics[width=\textwidth]{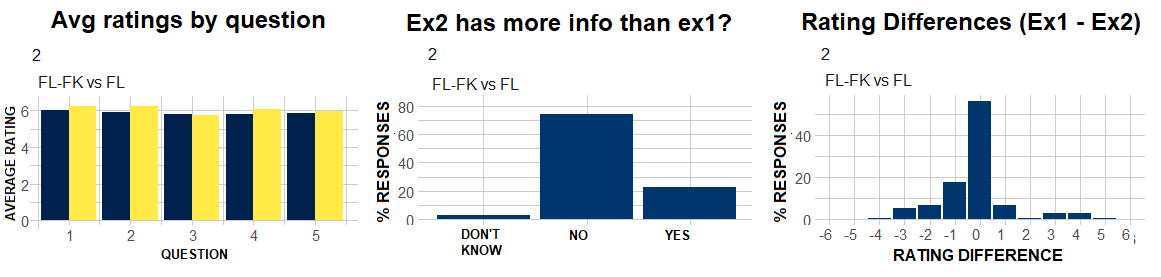}
    \caption{\emph{FL-FK} vs {FL}. The first graph shows the rating for each of the five questions outlined in Table \ref{tab:5Qs}. The centre graph shows whether participants found Explanation 2 to contain more information than Explanation 1. The right graph shows the difference between the ratings given to the two explanations.}
    \label{fig:FLFKvsCF}
\end{figure*}

\textbf{FL-FK vs FL} Recall that \emph{FK} removes background knowledge from explanations. 
We did not find any significant difference between the ratings given to  Explanation 1 (generated through \emph{FL-FK}) and Explanation 2 (generated through \emph{FL}). Around 75\% of the participants who rated these explanations stating that Explanation 2 did not contain any additional information. In a scenario on predicting heatwaves, the explanatory system added the following piece of information to Explanation 2 when not using the \emph{FK} rule:
\textit{35 degrees Celsius is greater than 24 and 25 degrees Celsius.}

Participants found that for heatwave prediction, the information that was added in Explanation 2 was obvious and common sense. For example, P33 stated that \textit{``Both explanations do a very good job of explaining why the heat wave determination was made.  The only additional info included in explanation 2 was common sense level information that wasn't required for the reader to understand why the prediction was made.''}

\begin{figure*}[t]
    \centering
    \includegraphics[width=\textwidth]{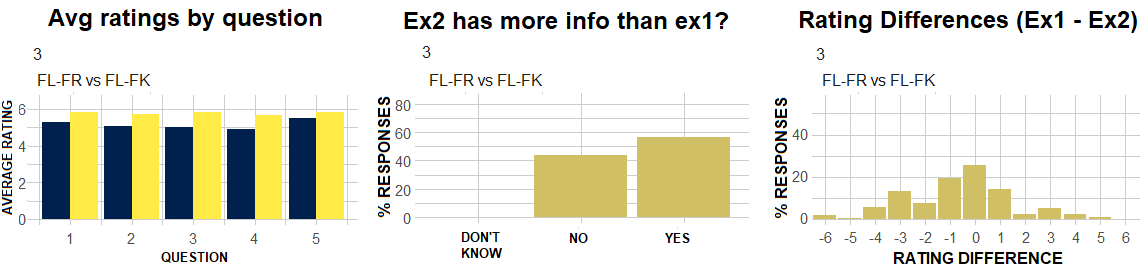}
    \caption{\emph{FL-FR} vs \emph{FL-FK}. The first graph shows the rating for each of the five questions outlined in Table \ref{tab:5Qs}. The centre graph shows whether participants found Explanation 2 to contain more information than Explanation 1. The right graph shows the difference between the ratings given to the two explanations.}
    \label{fig:CFRCvsCFFK}
\end{figure*}

\textbf{FL-FR vs FL-FK} For this pair, the results were somewhat split and dependent on the scenario and type of information. Responses from around 56\% of the participants indicated that not applying the \emph{FR} rule (leaving descriptions of the inference rules involved in reaching the conclusion in the explanation as footnotes) was more helpful in understanding and trusting the predictions. On the prediction of fog,  participants found these inference rules helpful in connecting pieces of information.  For example, P51 wrote:
    \textit{``The first one is disjointed without explaining what the information is leading up to. The second one includes the prediction of fog, and also says why these other conditions can cause fog.''}

Participants found some types of information, removed from an explanation when \emph{FK} is applied, helpful. On the prediction of damage due to a category 2 cyclone,  P62 mentioned:
    \textit{``Explanation one gave more details on how fast/damaging wind speeds might be when the storm makes landfall.''}

\subsection{Study 2 Results: Defence Experts}
\label{sec:results-study2}
Overall, we observed similar trends in terms of ratings; Explanation 2 was rated higher. Three of the experts rated Explanation 2 either slightly higher or the same as Explanation 1. Mostly, these experts rated their satisfaction with Explanation 2 a point or two higher than Explanation 1.  

One of the experts found the definitions (footnotes) useful in developing a comprehensive understanding of the situation. For example, Expert 1 stated:
    \textit{``The technical definition of `illegal fishing activities' that needn't involve any actual fish is not at all obvious and needed to make sense of the conclusion. Therefore the information about this rule provided by explanation 2 is genuinely additional and helpful.''}

For the pair \emph{FL vs no abstraction}, the experts had similar issues with Explanation 2; too much information. 
For example, Expert 5 wrote:
    \textit{``Explanation 2 repeated several facts, or made only trivial inference steps. I didn't see any content in explanation 2 that wasn't in explanation 1.''}

Experts agreed that not all of underlying reasoning information is relevant to people. MTurkers also indicated that it may be unnecessary to explain every reasoning or logical step taken to reach a conclusion. 
All experts echoed similar sentiments: we should avoid repeating information, avoid stating common sense knowledge, and avoid explaining every reasoning or logical step. However, the footnotes were generally helpful; it was apparent that most experts used the footnotes to confirm what they believed to be obvious.

%% file: 5-discussion.tex
\section{Discussion}
\label{sec:discussion}
There were two aims of the user evaluation: 1) to identify when to remove or leave different types of causal information in an explanation, and 1) to investigate when people prefer simple or complex explanations. 

\subsection{Explanation coherence matters}
In our model, using the \emph{Flatten Rules (FR)} abstraction step removed information (rule applications) that connected one or more inferred premises to a conclusion or intermediate conclusion. When the system did not use the \emph{FR} abstraction step, we placed the rule definitions as footnotes. 
These footnotes helped participants link causal information in the explanation together to understand the conclusions, that is, most of the participants commented that the presence of the rules was helpful. The rules in the footnotes assisted the explanation's coherency. It has been well established that coherency matters~\cite{thagard1978best,bovens2000coherentism,bovens2003solving,koslowski2008information,zemla2017evaluating}. We take coherency to mean how well the causal mechanisms of the explanation link together~\cite{zemla2017evaluating}. It is evident from the results in Figure~\ref{fig:avg_by_pairs_by_question_2022} that whether the \emph{FR} step was applied was one of the important reasons for the participants' ratings. In terms of our study, this means that we should not apply the \emph{FR} abstraction step. 

\subsection{Complexity preferences vs type of causal information}
Our system manipulated four types of causal information: told knowledge, background knowledge, inference rules, and inferred knowledge. While inference rules were found to be important, we observed conflicting views regarding background knowledge, which was removed from explanations by  \emph{Filter Knowledge (FK)}.  When comparing the pairs \emph{FL-FR-FK} and \emph{FL-FR}, many participants found information removed by  \emph{FK} was relevant in understanding the prediction.  P51 wrote for the  prediction of thunderstorm asthma:
    \textit{``The second one provides information on humidity levels that the first one was lacking. I can see why high humidity combined with lightning and pollen could cause an asthma attack.''}

All information the reasoning system uses is causally relevant to the predictions being made. However, people may not want or need all of that information. We found that participants rated explanations lower if the explanations contained causal information that was assumed to be common sense knowledge. This result is interesting because \citet{bechlivanidis2017concreteness} found that people did not penalise causally irrelevant information, information that does not exert any causal influence. This disparity could be due to extra information's role in the conclusion. In \citet{bechlivanidis2017concreteness}, information was completely irrelevant to the conclusion, while in our work, information was common sense but related to the conclusion.  Future works should investigate why people can ignore irrelevant information but react negatively to causally relevant common sense information.

\subsection{Participants preferred complex explanations} 
According to the opponent heuristic theory~\cite{johnson2019simplicity}, people prefer more complex explanations when asked to judge an explanation's likelihood or fit to the evidence. In our study,  participants were required to judge whether the evidence supported the prediction or not. Furthermore, according to  \citet{Lim2020}, people prefer simple explanations for simple events and complex ones for complex events. In our work, understanding and associating the explanations to the conclusions can be regarded as a complex task. We believe this could also explain why participants predominantly favoured the more complex explanations.

\citet{blanchard2020explanatory} and \citet{aronowitz2020experiential} provide various reasons for people seeking moderately abstract explanations, that is, explanations that contain the right level of detail. Such explanations aid people in doing mental simulations. We found that our participants preferred complex explanations over the simpler ones up to a certain point; the most complex explanation was not well received. Thus our results also support the claims in \citet{aronowitz2020experiential} in that explanations have to strike a balance between how much and what details to include in the explanation. 

We observed that the choice for simplicity or complexity was scenario and person-dependent. This suggests that while we observed a preference for complexity, \emph{whom} we are explaining to also matters: whether an explanation will be satisfactory to an individual depends on many factors \cite{lombrozo2017causal}.

\subsection{Limitations}
Our results should be read with a few caveats. We systematically generated three sets of pairs (rule combinations) such that each layer of an explanation was compared with the two layers below it, with and without the use of the \emph{Flatten Rules (FR)} step. Some rule combinations were not examined to manage the complexity of the study. However, we could have missed some relevant effects regarding the three abstraction rules or the types of causal information by not doing so.

We conducted our second study with only five experts, which is insufficient for statistical confidence in the results. Still, nonetheless, the qualitative insights provided by experts offer value beyond those of non-expert participants. We also analysed perceived preferences and not the actual utility of the explanations in helping make a decision. As noted by \citet{bechlivanidis2017concreteness}, preferences regarding explanations likely depend on one's aims; hence, situating the study in a task environment where people need to rely on the explanations would allow a different perspective.

%% file: 6-conclusion.tex
\section{Conclusion}
\label{sec:conclusion}
 Prior research has provided an understanding of when people prefer simple or complex explanations. We investigated the preference for simple or complex explanations when an artificial agent seeks to explain its conclusions to a human, by systematically varying the amount of causal information provided to the user. 
 Overall, we found a general preference for complex explanations. This finding aligns with existing research. At the same time, we note that the type of causal information determines the preference for complexity. 
 Because people's satisfaction with explanations depends on many contextual factors, it may not be so much whether people prefer simple or complex explanations, but \emph{what} types of causal information are present in the explanation and relevant to them. 
 

%% file: 7-appendix.tex
\appendices


\onecolumn
\section{Results of Friedman Test}
\label{sec:appendix-data-1}

\begin{center}
\begin{longtable}{p{1.5cm}|p{0.5cm}|p{3.0cm}|p{1cm}|p{1cm}|p{1cm}|p{1cm}|p{0.8cm}}
\caption{Details of the Friedman Test. There are eight columns: label of the pair, question number and text, average rating and standard deviation of explanation 1, average rating and standard deviation of explanation 2, $p$ value from conducting Friedman test, $\chi^2$ statistic, and $Kendall's~W$ or coefficient of concordance capturing the agreement among raters. $Kendall's~W$ was computed taking into consideration ratings for both explanations.} \label{tab:long} \\

\toprule 
    \multicolumn{1}{l|}{\textbf{Pair}} & 
    \multicolumn{1}{l|}{\textbf{Q\#}} & 
    \multicolumn{1}{l|}{\textbf{Question}} & 
    \thead{\textbf{Avg} \\ \textbf{Exp1} \\ \textbf{(SD)}} &
    \thead{\textbf{Avg} \\ \textbf{Exp2} \\ \textbf{(SD)}} &
    \multicolumn{1}{l|}{\textbf{p}} & 
    \multicolumn{1}{l|}{\textbf{$\chi^2$}} &
    \thead{\textbf{$Kendall's$} \\ \textbf{$W$}} \\
\midrule
\endfirsthead

\multicolumn{8}{c}%
{{\bfseries \tablename\ \thetable{} -- continued from previous page}} \\
\hline 
    \multicolumn{1}{l|}{\textbf{Pair}} & 
    \multicolumn{1}{l|}{\textbf{Q\#}} & 
    \multicolumn{1}{l|}{\textbf{Question}} & 
    \thead{\textbf{Avg} \\ \textbf{Exp1} \\ \textbf{(SD)}} &
    \thead{\textbf{Avg} \\ \textbf{Exp2} \\ \textbf{(SD)}} &
    \multicolumn{1}{l|}{\textbf{p}} & 
    \multicolumn{1}{l|}{\textbf{$\chi^2$}} &
    \thead{\textbf{$Kendall's$} \\ \textbf{$W$}} \\
\hline 
\endhead

 \bottomrule \multicolumn{3}{r}{{Continued on next page}} \\
\endfoot

\hline 
\endlastfoot

 FL-FR vs no abstraction  & 1 &  From the explanation, I understand why the given prediction has been made.  & 5.35 (1.4) & 5.92 (1.32) & $<0.001$ & 11.56 & 0.66\\
 FL-FR vs no abstraction  & 2 &  The explanation of why the prediction was made provides sufficient detail.  & 4.86 (1.78) & 5.97 (1.19) & $<0.001$ & 15.38 & 0.58\\
 FL-FR vs no abstraction  & 3 &  The explanation of why the prediction was made is satisfying.  & 4.92 (1.83) & 5.24 (1.85) & 0.083 & 3 & 0.44\\
 FL-FR vs no abstraction  & 4 &  The explanation of why the prediction was made is complete.  & 4.59 (1.89) & 5.92 (1.34) & $<0.001$ & 21.55 & 0.61\\
 FL-FR vs no abstraction  & 5 &  The explanation of why the prediction was made is trustworthy.  & 5.16 (1.48) &5.68 (1.49) & 0.003 & 8.91 & 0.53\\
 FL-FR vs FL  & 1 &  From the explanation, I understand why the given prediction has been made.  & 5.35 (1.59) &5.97 (1.34) & 0.005 & 8.07 & 0.63\\
 FL-FR vs FL  & 2 &  The explanation of why the prediction was made provides sufficient detail.  & 5.41 (1.54) &5.97 (1.53) & 0.003 & 9 & 0.64\\
 FL-FR vs FL  & 3 &  The explanation of why the prediction was made is satisfying.  & 5.09 (1.62) &5.74 (1.64) & 0.011 & 6.55 & 0.69\\
 FL-FR vs FL  & 4 &  The explanation of why the prediction was made is complete.  & 5.12 (1.72) &5.94 (1.41) & $<0.001$ & 12.8 & 0.64\\
 FL-FR vs FL  & 5 &  The explanation of why the prediction was made is trustworthy.  & 5.18 (1.66) &5.97 (1.4) & $<0.001$ & 14.22 & 0.68\\
 FL-FR-FK vs FL  & 1 &  From the explanation, I understand why the given prediction has been made.  & 5.34 (1.3) &6.44 (0.63) & $<0.001$ & 24.14 & 0.59\\
 FL-FR-FK vs FL  & 2 &  The explanation of why the prediction was made provides sufficient detail.  & 5.05 (1.43) &6.44 (0.67) & $<0.001$ & 27 & 0.57\\
 FL-FR-FK vs FL  & 3 &  The explanation of why the prediction was made is satisfying.  & 4.95 (1.56) &6.24 (1.04) & $<0.001$ & 20.16 & 0.63\\
 FL-FR-FK vs FL  & 4 &  The explanation of why the prediction was made is complete.  & 4.9 (1.58) &6.39 (0.74) & $<0.001$ & 27.13 & 0.63\\
 FL-FR-FK vs FL  & 5 &  The explanation of why the prediction was made is trustworthy.  & 5.27 (1.28) &6.22 (1.06) & $<0.001$ & 18.62 & 0.67\\
 FL-FK vs FL  & 1 &  From the explanation, I understand why the given prediction has been made.  & 6 (1.06) &6.23 (1.11) & 0.071 & 3.27 & 0.66\\
 FL-FK vs FL  & 2 &  The explanation of why the prediction was made provides sufficient detail.  & 5.91 (1.07) &6.26 (1.15) & 0.052 & 3.77 & 0.65\\
 FL-FK vs FL  & 3 &  The explanation of why the prediction was made is satisfying.  & 5.77 (1.26) &5.71 (1.6) & 0.513 & 0.43 & 0.58\\
 FL-FK vs FL  & 4 &  The explanation of why the prediction was made is complete.  & 5.8 (1.18) &6.09 (1.25) & 0.071 & 3.27 & 0.68\\
 FL-FK vs FL  & 5 &  The explanation of why the prediction was made is trustworthy.  & 5.83 (1.4) &5.94 (1.66) & 0.166 & 1.92 & 0.65\\
 FL-FR-FK vs FL-FK  & 1 &  From the explanation, I understand why the given prediction has been made.  & 5.15 (1.46) &6.15 (1.16) & $<0.001$ & 15.7 & 0.49\\
 FL-FR-FK vs FL-FK  & 2 &  The explanation of why the prediction was made provides sufficient detail.  & 4.85 (1.79) &6.1 (1.17) & $<0.001$ & 13.5 & 0.56\\
 FL-FR-FK vs FL-FK  & 3 &  The explanation of why the prediction was made is satisfying.  & 4.9 (1.86) &6.23 (0.96) & $<0.001$ & 13.5 & 0.45\\
 FL-FR-FK vs FL-FK  & 4 &  The explanation of why the prediction was made is complete.  & 4.51 (1.83) &6.15 (1.04) & $<0.001$ & 22.15 & 0.56\\
 FL-FR-FK vs FL-FK  & 5 &  The explanation of why the prediction was made is trustworthy.  & 5.1 (1.55) &6.15 (0.9) & $<0.001$ & 16.2 & 0.58\\
 FL-FR vs FL-FK  & 1 &  From the explanation, I understand why the given prediction has been made.  & 5.31 (1.47) &5.87 (1.22) & 0.05 & 3.85 & 0.44\\
 FL-FR vs FL-FK  & 2 &  The explanation of why the prediction was made provides sufficient detail.  & 5.1 (1.65) &5.72 (1.32) & 0.106 & 2.61 & 0.4\\
 FL-FR vs FL-FK  & 3 &  The explanation of why the prediction was made is satisfying.  & 5 (1.81) &5.82 (1.17) & 0.077 & 3.13 & 0.43\\
 FL-FR vs FL-FK  & 4 &  The explanation of why the prediction was made is complete.  & 4.9 (1.76) &5.69 (1.26) & 0.068 & 3.33 & 0.46\\
 FL-FR vs FL-FK  & 5 &  The explanation of why the prediction was made is trustworthy.  & 5.51 (1.37) &5.82 (1.17) & 0.239 & 1.38 & 0.5\\
 FL-FR-FK vs FL-FR  & 1 &  From the explanation, I understand why the given prediction has been made.  & 5.58 (0.78) &6.08 (0.76) & $<0.001$ & 17 & 0.7\\
 FL-FR-FK vs FL-FR  & 2 &  The explanation of why the prediction was made provides sufficient detail.  & 5.18 (0.96) &5.98 (0.97) & $<0.001$ & 19 & 0.58\\
 FL-FR-FK vs FL-FR  & 3 &  The explanation of why the prediction was made is satisfying.  & 5.05 (0.96) &5.83 (0.93) & $<0.001$ & 21 & 0.64\\
 FL-FR-FK vs FL-FR  & 4 &  The explanation of why the prediction was made is complete.  & 4.78 (1.17) &5.75 (1.13) & $<0.001$ & 24 & 0.67\\
 FL-FR-FK vs FL-FR  & 5 &  The explanation of why the prediction was made is trustworthy.  & 5.7 (0.91) &6.1 (0.98) & 0.002 & 9.31 & 0.68\\
 FL vs no abstraction  & 1 &  From the explanation, I understand why the given prediction has been made.  & 5.73 (1.22) &5.84 (1.26) & 0.439 & 0.6 & 0.57\\
 FL vs no abstraction  & 2 &  The explanation of why the prediction was made provides sufficient detail.  & 5.51 (1.35) &5.89 (1.15) & 0.018 & 5.56 & 0.65\\
 FL vs no abstraction  & 3 &  The explanation of why the prediction was made is satisfying.  & 5.57 (1.54) &5.14 (1.92) & 0.67 & 0.18 & 0.51\\
 FL vs no abstraction  & 4 &  The explanation of why the prediction was made is complete.  & 5.35 (1.48) &5.89 (1.02) & 0.018 & 5.56 & 0.58\\
 FL vs no abstraction  & 5 &  The explanation of why the prediction was made is trustworthy.  & 5.7 (1.43) &5.46 (1.52) & 0.637 & 0.22 & 0.56\\
 FL-FK vs no abstraction  & 1 &  From the explanation, I understand why the given prediction has been made.  & 5.55 (1.39) &6.15 (1.28) & 0.003 & 8.89 & 0.66\\
 FL-FK vs no abstraction  & 2 &  The explanation of why the prediction was made provides sufficient detail.  & 5.12 (1.56) &6.45 (0.67) & $<0.001$ & 13.76 & 0.5\\
 FL-FK vs no abstraction  & 3 &  The explanation of why the prediction was made is satisfying.  & 4.94 (1.6) &5.82 (1.51) & 0.012 & 6.26 & 0.38\\
 FL-FK vs no abstraction  & 4 &  The explanation of why the prediction was made is complete.  & 4.7 (1.93) &6.39 (0.75) & $<0.001$ & 15.7 & 0.47\\
 FL-FK vs no abstraction  & 5 &  The explanation of why the prediction was made is trustworthy.  & 5.24 (1.64) &6.21 (1.14) & $<0.001$ & 12.8 & 0.58\\

\end{longtable}
\end{center}

\newpage
\section{Page Descriptions}
\label{sec:appendix-scenario-descs}
The following provides the pages presented to participants in our studies. There are 18 pages divided into 2 groups of 9 pages.



\section*{Supplementary material}

**GROUP 1 PAGES**

**Prediction: A blizzard is likely to develop in Alaska on the 15th of January 2019**

**Pair/Setting: FL-FR vs no abstraction / 1**

| | |
|---|---|
| **Explanation 1:** The National Weather Service has forecast cold ground temperatures, high humidity, high winds, and heavy snowfall for Alaska on the 15th of January 2019. | **Explanation 2:** The National Weather Service has forecast cold ground temperatures in Alaska on the 15th of January 2019. Consequently, cold ground temperatures have been forecast for Alaska on the 15th of January 2019. The National Weather Service has forecast high humidity for Alaska on the 15th of January 2019. Consequently, high humidity levels have been forecast for Alaska on the 15th of January 2019. The National Weather Service has forecast high winds for Alaska on the 15th of January 2019. Consequently, high winds have been forecast for Alaska on the 15th of January 2019. The National Weather Service has forecast heavy snowfall for Alaska on the 15th of January 2019. Consequently, heavy snow falls have been forecast for Alaska on the 15th of January 2019. Consequently, all of the following hold: high winds have been forecast for Alaska on the 15th of January 2019; and heavy snow falls have been forecast for Alaska on the 15th of January 2019. Consequently, all of the following hold: high humidity levels have been forecast for Alaska on the 15th of January 2019; high winds have been forecast for Alaska on the 15th of January 2019; and heavy snow falls have been forecast for Alaska on the 15th of January 2019. Consequently, all of the following hold: cold ground temperatures have been forecast for Alska on the 15th of January 2019; high humidity levels have been forecast for Alaska on the 15th of January 2019; high winds have been forecast for Alaska on the 15th of January 2019; and heavy snow falls have been forecast for Alaska on the 15th of January 2019. It follows that a blizzard is likely to develop in Alaska on the 15th of January 2019[1]. |
| | 1: A blizzard is likely to occur when cold ground temperatures, high humidity, strong winds, and heavy snowfalls are present. |

**Prediction: The Sea Witch (a ship) is engaging in illegal fishing on Sunday the 3rd of June 2018.**
**Pair/Setting: FL-FR vs FL / 1**

**Explanation 1:** The Maritime Authority indicates that the Sea Witch is not authorised to transfer fish. Transshipment of a catch at sea between two vessels is indicated when: more than twenty AIS signals, that are within 500 m of each other, are detected from the two vessels over 3 hours; the two vessels are travelling at less than 2 knots (are stationary); one of the vessels is a refrigerated cargo ship; and the two vessels are located more than 40 km from shore. The Sea Witch is engaged in transshipment with the Fox Carrier on Sunday the 3rd of June 2018 at 7:43:00 AM at 160.32N 25.52W.

**Explanation 2:** The Maritime Authority indicates that the Sea Witch is not authorised to transfer fish. Transshipment of a catch at sea between two vessels is indicated when: more than twenty AIS signals, that are within 500 m of each other, are detected from the two vessels over 3 hours; the two vessels are travelling at less than 2 knots (are stationary); one of the vessels is a refrigerated cargo ship, and the two vessels are located more than 40 km from shore. The Sea Witch is engaged in transshipment with the Fox Carrier on Sunday the 3rd of June 2018 at 7:43:00 AM at 160.32N 25.52W. Consequently, there exists a vessel with which the Sea Witch is engaging in transshipment on Sunday the 3rd of June 2018 at 7:43:00 AM. It follows that the vessel is engaging in illegal fishing activities[1].

1: A vessel is engaging in illegal fishing activities if it is not authorised to transfer fish, but is engaging in transshipment with another vessel.

**Prediction:** The Sea Witch (a ship) is engaging in illegal fishing activities on Sunday the 3rd of June 2018.
**Pair/Setting:** FL-FR-FK vs FL / 1

**Explanation 1:** The Maritime Authority indicates that the Sea Witch is not authorised to transfer fish. Aerial Patrol reported the presence of the Sea Witch at 160.32N 25.52W on Sunday the 3rd of June 2018 at 7:43:00 AM. Aerial Patrol reported the presence of the Fox Carrier at 160.32N 25.521W on Sunday the 3rd of June 2018 at 7:43:00 AM. Consequently, the Sea Witch is colocated with Fox Carrier at this time. Imagery provided by the Aerial Patrol indicated that the Sea Witch had turtles on board. Consequently, the Sea Witch is engaged in transshipment with the Fox Carrier on Sunday the 3rd of June 2018 at 7:43:00 AM at 160.32N 25.52W.

**Explanation 2:** The Maritime Authority indicates that the Sea Witch is not authorised to transfer fish. Aerial Patrol reported the presence of the Sea Witch at 160.32N 25.52W on Sunday the 3rd of June 2018 at 7:43:00 AM. Aerial Patrol reported the presence of the Fox Carrier at 160.32N 25.521W on Sunday the 3rd of June 2018 at 7:43:00 AM. The coordinates 160.32N 25.52W and 160.32N 25.521W are less than 500m apart. AIS track data indicates that the Sea Witch has been stationary for more than 3 hours. AIS track data indicates that the Fox Carrier has been stationary for more than 3 hours. Consequently, the Sea Witch is colocated with the Fox Carrier on Sunday the 3rd of June 2018 at 7:43:00 AM[1]. The Maritime Authority says that the Fox Carrier is a refrigerated cargo ship. Imagery provided by the Aerial Patrol indicated that the Sea Witch had turtles on board. Consequently, there exists a species that is on board the Sea Witch on Sunday the 3rd of June 2018 at 7:43:00 AM. The encounter is taking place more than 40 km from shore. Consequently, the Sea Witch is engaged in transshipment with the Fox Carrier on Sunday the 3rd of June 2018 at 7:43:00 AM at 160.32N 25.52W[2]. Consequently, there exists a vessel with which the Sea Witch is engaging in transshipment on Sunday the 3rd of June 2018 at 7:43:00 AM. It follows that the vessel is engaging in illegal fishing activities[3].

1: Two vessels are colocated if they are stationary and have been positioned within 500m of each other for over 3 hours.
2: A vessel is engaging in transshipment with a refrigerated cargo ship if the two vessels are colocated, the encounter is occurring more than 40 km from shore, and there is a species on board the vessel.
3: A vessel is engaging in illegal fishing activities if it is not authorised to transfer fish but is engaging in transshipment with another vessel.

**Prediction: A heatwave is imminent on Tuesday the 1st of December 2020, in Melbourne, Australia.**
**Pair/Setting: Heatwave FL-FK vs FL / 2**

**Explanation 1:** The Bureau of Meteorology has forecast maximum temperatures of over 35 degrees Celsius between Tuesday the 1st of December and Thursday the 3rd of December 2020 in Melbourne, Australia. Over the past thirty days, the average maximum temperature in Melbourne, Australia, has been 25 degrees Celsius. The Bureau of Meteorology indicates that the average maximum temperature for Melbourne, Australia, in December is 24 degrees Celsius. It follows that a heatwave is imminent on Tuesday the 1st of December 2020, in Melbourne, Australia[1].

1: A heatwave is imminent if the maximum temperatures forecast over the next three days are (i) significantly higher than the average temperatures over the past thirty days, and (ii) are higher than what is typical for the time of year.

**Explanation 2:** The Bureau of Meteorology has forecast maximum temperatures of over 35 degrees Celsius between Tuesday the 1st of December and Thursday the 3rd of December 2020 in Melbourne, Australia. Over the past thirty days, the average maximum temperature in Melbourne, Australia, has been 25 degrees Celsius. The Bureau of Meteorology indicates that the average maximum temperature for Melbourne, Australia, in December is 24 degrees Celsius. 35 degrees Celsius is greater than 24 and 25 degrees Celsius. It follows that a heatwave is imminent on Tuesday the 1st of December 2020, in Melbourne, Australia[1].

1: A heatwave is imminent if the maximum temperatures forecast over the next three days are (i) significantly higher than the average temperatures over the past thirty days, and (ii) are higher than what is typical for the time of year.

**Prediction: A moderate level of cyclone damage to coastal cities in the Northern Territory is likely on the 16th of March 2018.**
**Pair/Setting: Cyclone-costal FL-FR-FK vs FL-FK / 3**

| **Explanation 1:** The Bureau of Meteorology has detected a tropical cyclone in the western Arafura Sea on the 15th of March 2018. A cyclone with a wind speed of 130 km/hr is classified as a Category 2 cyclone. Consequently, a moderate level of cyclone damage to coastal cities in the Northern Territory is likely on the 16th of March 2018. | **Explanation 2:** The Bureau of Meterorology has detected a tropical cyclone in the western Arafura Sea on the 15th of March 2018. A cyclone with a wind speed of 130 km/hr is classified as a Category 2 cyclone. Consequently, a moderate level of cyclone damage to coastal cities in the Northern Territory is likely on the 16th of March 20181. 1 A category 2 cyclone is likely to cause moderate damage to homes |
|---|---|

**Prediction: Fog is likely on Thursday the 3rd of December 2020, in Melbourne, Australia.**
**Pair/Setting: Fog FL-FR (s1) vs FL-FK (s2) / 3**

| **Explanation 1:** The Bureau of Meteorology has forecast clear skies overnight leading up to Thursday 3rd of December 2020 in Melbourne, Australia. The Bureau of Meteorology has forecast high moisture content in the air, and light winds, on Thursday 3rd of December 2020, in Melbourne Australia. | **Explanation 2:** The Bureau of Meteorology has forecast high moisture content in the air, and light winds, for Thursday 3rd of December 2020, in Melbourne Australia. It follows that fog is likely on Thursday 3rd of December 2020 in Melbourne, Australia1.  1 Fog is likely to form in the presence of light winds and high moisture content in the air, after a night of clear skies. |
|---|---|

**Prediction: There is a high risk of epidemic thunderstorm asthma on Tuesday the 1st of September 2020.**

**Pair/Setting: Thunderstorm Asthma FL-FR-FK vs FL-FR / 1**

| **Explanation 1:** The Bureau of Meteorology indicates that grass pollen levels will be high on Tuesday the 1st of September 2020. A thunderstorm with lightning has been forecast for Tuesday the 1st of September 2020 by the Bureau of Meteorology. Consequently, there is a high risk of epidemic thunderstorm asthma on Tuesday the 1st of September 2020. | **Explanation 2:** The Bureau of Meteorology indicates that grass pollen levels will be high on Tuesday the 1st of September 2020. The Bureau of Meteorology has forecast high humidity for Tuesday the 1st of September 2020. A thunderstorm with lightning has been forecast for Tuesday the 1st of September 2020 by the Bureau of Meteorology. Consequently, there is a high risk of epidemic thunderstorm asthma on Tuesday the 1st of September 2020. |
|---|---|

**Prediction: The Sea Witch (a ship) is engaging in illegal fishing activities in the Norfolk Island Offshore Fishery on Sunday the 3rd of June 2018 at 7:43:00 AM**

**Pair/Setting: Sea Witch (False MMSI) FL vs no abstraction / 1**

| **Explanation 1:** The Maritime Authority has indicated that the Sea Witch is emitting a false Maritime Mobile Service Identity (MMSI) number on Sunday the 3rd of June 2018 at 7:43:00 AM. Aerial Patrol reported the presence of the Sea Witch in the Norfolk Island Offshore Fishery on Sunday the 3rd of June 2018 at 7:43:00 AM. The Norfolk Island Offshore Fishery is an exclusive economic zone. It follows that the vessel is likely to be engaging in illegal fishing activities[1].<br><br>1: Analysis of historical data indicates that a vessel emitting a false MMSI number in an exclusive economic zone is likely to be engaging in illegal fishing activities. | **Explanation 2:** The Maritime Authority has indicated that the Sea Witch is emitting a false Maritime Mobile Service Identity (MMSI) number on Sunday the 3rd of June 2018 at 7:43:00 AM. Consequently, the Sea Witch is emitting a false MMSI number on Sunday the 3rd of June 2018 at 7:43:00 AM. Aerial Patrol reported the presence of the Sea Witch in the Norfolk Island Offshore Fishery on Sunday the 3rd of June 2018 at 7:43:00 AM. Consequently, the Sea Witch is located at Norfolk Island Offshore Fishery on Sunday the 3rd of June 2018 at 7:43:00 AM. The Norfolk Island Offshore Fishery is an exclusive economic zone. Consequently, the Norfolk Island Offshore Fishery is an exclusive economic zone. Consequently, all of the following hold: the Sea Witch is located at Norfolk Island Offshore Fishery on Sunday the 3rd of June 2018 at 7:43:00 AM; and the Norfolk Island Offshore Fishery is an exclusive economic zone. Consequently, all of the following hold: the Sea Witch is emitting a false MMSI number on Sunday the 3rd of June 2018 at 7:43:00 AM; the Sea Witch is located at Norfolk Island Offshore Fishery on Sunday the 3rd of June 2018 at 7:43:00 AM; and the Norfolk Island Offshore Fishery is an exclusive economic zone. It follows that the vessel is likely to be engaging in illegal fishing activities[1].<br><br>1: Analysis of historical data indicates that a vessel emitting a false MMSI number in an exclusive economic zone is likely to be engaging in illegal fishing activities. |
|---|---|

**Prediction: An eruption of the Alopa volcano is imminent on the 10th of April 2019.**
**Pair/Setting: Volcano FL-FK vs no abstraction / 2**

**Explanation 1:** The Bureau of Meteorology has detected an increasing frequency of earthquakes beneath the Alopa volcano on the 10th of April 2019. Increased steaming activity and magma intrusions have also been observed. It follows that an eruption is imminent[1].

1: Increasing frequency of earthquakes, magma intrusions, and increased steaming activity, indicates that an eruption is imminent.

**Explanation 2:** The Bureau of Meteorology (BOM) has detected an increasing frequency of earthquakes beneath the Alopa volcano on the 10th of April 2019. Consequently, an increasing frequency of earthquakes has been detected beneath the Alopa volcano on the 10th of April 2019. Local authorities have reported ground swelling and fissuring. Consequently, ground swelling and fissuring have been observed at the Alopa volcano on the 10th of April 2019. Consequently, magma intrusions are present[1]. Local authorities have also reported gray steam emission at the site. Consequently, gray steam emission has been observed at the Alopa volcano on the 10th of April 2019. Consequently, increased steaming activity has been observed[2]. Consequently, all of the following hold: magma intrusions are present at the Alopa volcano on the 10th of April 2019; increased steaming activity has been observed at the Alopa volcano on the 10th of April 2019; and an increasing frequency of earthquakes has been detected beneath the Alopa volcano on the 10th of April 2019. It follows that an eruption is imminent[3].

1: Ground swelling indicates the presence of magma intrusions.
2: A change in the colour of emitted steam from white to gray, due to the presence of ash, indicates increased steaming activity.
3: Increasing frequency of earthquakes, magma intrusions, and increased steaming activity, indicates that an eruption is imminent.

**GROUP 2 PAGES**

**Prediction: The Sea Witch (a ship) is engaging in illegal fishing activities on Sunday the 3rd of June 2018.**
**Pair/Setting: Transshipment 1 FL-FR vs FL / 1**

**Explanation 1:** The Maritime Authority indicates that the Sea Witch is not authorised to transfer fish. Aerial Patrol reported the presence of the Sea Witch at 160.32N 25.52W on Sunday the 3rd of June 2018 at 7:43:00 AM. Aerial Patrol reported the presence of the Fox Carrier at 160.32N 25.521W on Sunday the 3rd of June 2018 at 7:43:00 AM. The coordinates 160.32N 25.52W and 160.32N 25.521W are less than 500m apart. AIS track data indicates that the Sea Witch has been stationary for more than 3 hours. AIS track data indicates that the Fox Carrier has been stationary for more than 3 hours. Consequently, the Sea Witch is colocated with the Fox Carrier on Sunday the 3rd of June 2018 at 7:43:00 AM. The Maritime Authority says that the Fox Carrier is a refrigerated cargo ship. Imagery provided by the Aerial Patrol indicated that the Sea Witch had turtles on board. The encounter is taking place more than 40km from shore. Consequently, the Sea Witch is engaged in transshipment with the Fox Carrier on Sunday the 3rd of June 2018 at 7:43:00 AM at 160.32N 25.52W.

**Explanation 2:** The Maritime Authority indicates that the Sea Witch is not authorised to transfer fish. Aerial Patrol reported the presence of the Sea Witch at 160.32N 25.52W on Sunday the 3rd of June 2018 at 7:43:00 AM. Aerial Patrol reported the presence of the Fox Carrier at 160.32N 25.521W on Sunday the 3rd of June 2018 at 7:43:00 AM. The coordinates 160.32N 25.52W and 160.32N 25.521W are less than 500m apart. AIS track data indicates that the Sea Witch has been stationary for more than 3 hours. AIS track data indicates that the Fox Carrier has been stationary for more than 3 hours. Consequently, the Sea Witch is colocated with the Fox Carrier on Sunday the 3rd of June 2018 at 7:43:00 AM[1]. The Maritime Authority says that the Fox Carrier is a refrigerated cargo ship. Imagery provided by the Aerial Patrol indicated that the Sea Witch had turtles on board. Consequently, there exists a species that is on board the Sea Witch on Sunday the 3rd of June 2018 at 7:43:00 AM. The encounter is taking place more than 40 km from shore. Consequently, the Sea Witch is engaged in transshipment with the Fox Carrier on Sunday the 3rd of June 2018 at 7:43:00 AM at 160.32N 25.52W[2]. Consequently, there exists a vessel with which the Sea Witch is engaging in transshipment on Sunday the 3rd of June 2018 at 7:43:00 AM. It follows that the vessel is engaging in illegal fishing activities[3].

1: Two vessels are colocated if they are stationary and have been positioned within 500m of each other for over 3 hours.
2: A vessel is engaging in transshipment with a refrigerated cargo ship if the two vessels are colocated, the encounter is occurring more than 40 km from shore, and there is a species on board the vessel.
3: A vessel is engaging in illegal fishing activities if it is not authorised to transfer fish but is engaging in transshipment with another vessel.

**Prediction: The Sea Witch (a ship) is engaging in illegal fishing on Sunday the 3rd of June 2018.**
**Pair/Setting: Transshipment 2 FL-FR vs no abstraction / 1**

| | |
|---|---|
| **Explanation 1:** The Maritime Authority indicates that the Sea Witch is not authorised to transfer fish. Transshipment of a catch at sea between two vessels is indicated when: more than twenty AIS signals, that are within 500 m of each other, are detected from the two vessels over 3 hours; the two vessels are travelling at less than 2 knots (are stationary); one of the vessels is a refrigerated cargo ship; and the two vessels are located more than 40 km from shore. The Sea Witch is engaged in transshipment with the Fox Carrier on Sunday the 3rd of June 2018 at 7:43:00 AM at 160.32N 25.52W. | **Explanation 2:** The Maritime Authority indicates that the Sea Witch is not authorised to transfer fish. Consequently, the Sea Witch is not permitted to transfer fish. Transshipment of a catch at sea between two vessels is indicated when: more than twenty AIS signals, that are within 500 m of each other, are detected from the two vessels over 3 hours; the two vessels are travelling at less than 2 knots (are stationary); one of the vessels is a refrigerated cargo ship, and the two vessels are located more than 40 km from shore. The Sea Witch is engaged in transshipment with the Fox Carrier on Sunday the 3rd of June 2018 at 7:43:00 AM at 160.32N 25.52W. Consequently, there exists a vessel with which the Sea Witch is engaging in transshipment on Sunday the 3rd of June 2018 at 7:43:00 AM. Consequently, all of the following hold: the Sea Witch is not permitted to transfer fish; and there exists a vessel with which the Sea Witch is engaging in transshipment on Sunday the 3rd of June 2018 at 7:43:00 AM. It follows that the vessel is engaging in illegal fishing activities[1]. <br><br> 1: A vessel is engaging in illegal fishing activities if it is not authorised to transfer fish, but is engaging in transshipment with another vessel. |

**Prediction: Paula Lands is in the Celtic Sea on the 26th of January 2013.**
**Pair/Setting: What Region is She In? FL-FK vs FL / 2**

| | |
|---|---|
| **Explanation 1:** Paula Lands said that she was on a yacht at 56.8075N 26.8753W on the 26th of January 2013. Consequently, Paula Lands was located at 56.8075N 26.8753W on the 26th of January 2013[1]. The coordinate 56.8075N 26.8753W is in the Celtic Sea region.<br><br>1: An entity is 'at' a coordinate if they are 'directly attached' to that coordinate. | **Explanation 2:** Paula Lands is female and animate (alive). The Celtic Sea is an extended surface. Consequently, the Celtic Sea is inanimate (not a person or alive)[1]. Paula Lands said that she was on a yacht at 56.8075N 26.8753W on the 26th of January 2013. Consequently, Paula Lands was located at 56.8075N 26.8753W on the 26th of January 2013[2]. The coordinate 56.8075N 26.8753W is in the polygon representing the Celtic Sea. Consequently, the coordinate 56.8075N 26.8753W is in the Celtic Sea region[3].<br><br>1: An entity is inanimate if it is an extended surface.<br>2: An entity X is 'at' a coordinate if they are 'directly attached' to that coordinate.<br>3: The coordinate C is in the region Y if Y is an extended surface and C is in the polygon defined by that surface. |

**Prediction: A moderate level of cyclone damage to coastal cities in the Northern Territory is likely on the 16th of March 2018.**
**Pair/Setting: Cyclone-costal FL-FR vs FL-FK / 3**

| | |
|---|---|
| **Explanation 1:** The Bureau of Meteorology has detected a tropical cyclone in the western Arafura Sea on the 15th of March 2018. The Bureau of Meteorology estimates that the Arafura Sea cyclone will reach landfall with a wind speed of approximately 130 km/hr on the 16th of March 2018. A cyclone with a wind speed of 130 km/hr is classified as a Category 2 cyclone. Consequently, a moderate level of cyclone damage to coastal cities in the Northern Territory is likely on the 16th of March 2018. | **Explanation 2:** The Bureau of Meteorology has detected a tropical cyclone in the western Arafura Sea on the 15th of March 2018. A cyclone with a wind speed of 130 km/hr is classified as a Category 2 cyclone. Consequently, a moderate level of cyclone damage to coastal cities in the Northern Territory is likely on the 16th of March 2018[1].<br><br>1 A category 2 cyclone is likely to cause moderate damage to homes |

**Prediction: An eruption of the Alopa volcano is imminent on the 10th of April 2019.**
**Pair/Setting: Volcano FL-FR-FK vs FL-FR / 1**

| Explanation 1: The Bureau of Meteorology (BOM) has detected an increasing frequency of earthquakes beneath the Alopa volcano on the 10th of April 2019. Increased steaming activity has been observed, and magma intrusions are present. | Explanation 2: The Bureau of Meteorology (BOM) has detected an increasing frequency of earthquakes beneath the Alopa volcano on the 10th of April 2019. Local authorities have reported ground swelling and fissuring. Consequently, magma intrusions are present. Local authorities have also reported gray steam emission at the site. Consequently, increased steaming activity has been observed. |
|---|---|

**Prediction: There is a high risk of epidemic thunderstorm asthma on Tuesday the 1st of September 2020.**
**Pair/Setting: Thunderstorm Asthma FL-FR-FK vs FL-FK / 3**

| Explanation 1: The Bureau of Meteorology indicates that grass pollen levels will be high on Tuesday the 1st of September 2020. A thunderstorm with lightning has been forecast for Tuesday the 1st of September 2020 by the Bureau of Meteorology. Consequently, there is a high risk of epidemic thunderstorm asthma on Tuesday the 1st of September 2020. | Explanation 2: The Bureau of Meteorology indicates that grass pollen levels will be high on Tuesday the 1st of September 2020. A thunderstorm with lightning has also been forecast. It follows that the risk of epidemic thunderstorm asthma is high1.     1 The risk of epidemic thunderstorm asthma is high when a thunderstorm, with lightning, is forecast at a time of high grass pollen levels and high humidity. |
|---|---|

**Prediction: The Sea Witch (a ship) is likely engaging in illegal fishing activities in the Norfolk Island Offshore Fishery on Sunday the 3rd of June 2018 at 7:43:00 AM.**
**Pair/Setting: Turtles FL-FR-FK vs FL / 1**

| | |
|---|---|
| **Explanation 1:** Aerial Patrol reported the presence of the Sea Witch at 156.12N 20.4W, about 2 nautical miles north of Norfolk Island on Sunday the 3rd of June 2018 at 7:43:00 AM. Consequently, the Sea Witch is operating in the Norfolk Island Offshore Fishery exclusive economic zone (EEZ). The Maritime Authority indicates that the Sea Witch is not licenced to catch turtles in the Norfolk Island Offshore Fishery. Imagery provided by the Aerial Patrol indicated that the Sea Witch had turtles on board. Consequently, there exists a species on board the Sea Witch that it is not permitted to catch. | **Explanation 2:** Aerial Patrol reported the presence of the Sea Witch at 156.12N 20.4W, about 2 nautical miles north of Norfolk Island on Sunday the 3rd of June 2018 at 7:43:00 AM. The Norfolk Island Offshore Fishery is a water surface. The coordinate 156.12N 20.4W is in the polygon representing the Norfolk Island Offshore Fishery. Consequently, the Sea Witch is operating in the Norfolk Island Offshore Fishery exclusive economic zone (EEZ)[1]. The Maritime Authority indicates that the Sea Witch is not licenced to catch turtles in the Norfolk Island Offshore Fishery. Imagery provided by the Aerial Patrol indicated that the Sea Witch had turtles on board. Consequently, there exists a species on board the Sea Witch that it is not permitted to catch. It follows that the vessel is engaging in illegal fishing activities[2].<br><br>1: A vessel is operating in a zone if that zone is a water surface, and the vessel is located at a coordinate within the polygon defined by that surface.<br>2: A vessel is engaging in illegal fishing activities in a zone if it is operating in that zone, and there exists a species on board the vessel that it is not permitted to catch. |

**Prediction: A blizzard is likely to develop in Alaska on the 15th of January 2019**
**Pair/Setting: Blizzard FL vs no abstraction / 1**

| | |
|---|---|
| **Explanation 1:** The National Weather Service has forecast cold ground temperatures, high humidity, high winds, and heavy snowfall for Alaska on the 15th of January 2019. It follows that a blizzard is likely to develop in Alaska on the 15th of January 2019[1].<br>1: A blizzard is likely to occur when cold ground temperatures, high humidity, strong winds, and heavy snowfalls are present. | **Explanation 2:** The National Weather Service has forecast cold ground temperatures in Alaska on the 15th of January 2019. Consequently, cold ground temperatures have been forecast for Alaska on the 15th of January 2019. The National Weather Service has forecast high humidity for Alaska on the 15th of January 2019. Consequently, high humidity levels have been forecast for Alaska on the 15th of January 2019. The National Weather Service has forecast high winds for Alaska on the 15th of January 2019. Consequently, high winds have been forecast for Alaska on the 15th of January 2019. The National Weather Service has forecast heavy snowfall for Alaska on the 15th of January 2019. Consequently, heavy snow falls have been forecast for Alaska on the 15th of January 2019. Consequently, all of the following hold: high winds have been forecast for Alaska on the 15th of January 2019; heavy snow falls have been forecast for Alaska on the 15th of January 2019. Consequently, all of the following hold: high humidity levels have been forecast for Alaska on the 15th of January 2019; high winds have been forecast for Alaska on the 15th of January 2019; heavy snow falls have been forecast for Alaska on the 15th of January 2019. Consequently, all of the following hold: cold ground temperatures have been forecast for Alska on the 15th of January 2019; high humidity levels have been forecast for Alaska on the 15th of January 2019; high winds have been forecast for Alaska on the 15th of January 2019; heavy snow falls have been forecast for Alaska on the 15th of January 2019. It follows that a blizzard is likely to develop in Alaska on the 15th of January 2019[1].<br><br>1 A blizzard is likely to occur when cold ground temperatures, high humidity, strong winds, and heavy snowfalls are present. |

**Prediction: The West Freighter (a ship) is engaging in illegal dumping in the Norfolk Island Offshore Fishery on Monday the 4th of June 2019 at 11:00 AM.**
**Pair/Setting: West Freighter Dumping FL-FK vs no abstraction / 2**

**Explanation 1:** Imagery provided by the Aerial Patrol shows an oil slick trailing from the West Freighter on Monday the 4th of June 2018 at 11:00 AM. Aerial Patrol reported the presence of the West Freighter at 156.12N 20.4W, about 2 nautical miles north of Norfolk Island at that time. Consequently, the West Freighter is operating in the Norfolk Island Offshore Fishery exclusive economic zone (EEZ) on Monday the 4th of June 2018 at 11:00 AM[1]. It follows that the West Freighter is engaging in illegal dumping in the Norfolk Island Offshore Fishery on Monday the 4th of June 2018 at 11:00 AM[2].

1: A vessel is operating in a zone if that zone is a water surface, and the vessel is located at a coordinate within the polygon defined by that surface.
2: A vessel is engaging in illegal dumping activities in a zone if it is operating in that zone, and dumped material can be seen trailing the vessel.

**Explanation 2:** Aerial Patrol reported the presence of the West Freighter at 156.12N 20.4W, about 2 nautical miles north of Norfolk Island on Monday the 4th of June 2018 at 11:00 AM. Consequently, the West Freighter is located at 156.12N 20.4W on Monday the 4th of June 2018 at 11:00 AM. The Norfolk Island Offshore Fishery is a water surface. Consequently, the Norfolk Island Offshore Fishery is a water surface. The coordinate 156.12N 20.4W is in the polygon representing the Norfolk Island Offshore Fishery. Consequently, all of the following hold: the Norfolk Island Offshore Fishery is a water surface; and the coordinate 156.12N 20.4W is in the polygon representing the Norfolk Island Offshore Fishery. Consequently, all of the following hold: the West Freighter is located at 156.12N 20.4W on Monday the 4th of June 2018 at 11:00 AM; the Norfolk Island Offshore Fishery is a water surface; and the coordinate 156.12N 20.4W is in the polygon representing the Norfolk Island Offshore Fishery. Consequently, the West Freighter is operating in the Norfolk Island Offshore Fishery exclusive economic zone (EEZ) on Monday the 4th of June 2018 at 11:00 AM[1]. Imagery provided by the Aerial Patrol shows an oil slick trailing from the West Freighter on Monday the 4th of June 2018 at 11:00 AM. Consequently, an oil slick is trailing the West Freighter on Monday the 4th of June 2018 at 11:00 AM. Consequently, all of the following hold: the West Freighter is operating in the Norfolk Island Offshore Fishery exclusive economic zone (EEZ) on Monday the 4th of June 2018 at 11:00 AM; and an oil slick is trailing the West Freighter on Monday the 4th of June 2018 at 11:00 AM. It follows that the West Freighter is engaging in illegal dumping in the Norfolk Island Offshore Fishery on Monday the 4th of June 2018 at 11:00 AM[2].

1: A vessel is operating in a zone if that zone is a water surface, and the vessel is located at a coordinate within the polygon defined by that surface.

2: A vessel is engaging in illegal dumping activities in a zone if it is operating in that zone, and dumped material can be seen trailing the vessel.

